\documentclass[12pt]{article}
\usepackage{graphicx}
\usepackage{amsmath}
\usepackage{longtable}

\begin{document}

% 4 rows Astronomy Letters style:
\makeatletter
\renewcommand*{\@cite}[2]{{#2}}
\renewcommand*{\@biblabel}[1]{#1.\hfill}
\makeatother

\title{The Red Giant Branch in the Tycho-2 Catalogue}
\author{G.~A.~Gontcharov\thanks{E-mail: georgegontcharov@yahoo.com}}

\maketitle

Pulkovo Astronomical Observatory, Russian Academy of Sciences, Pul\-kov\-skoe sh. 65, St. Petersburg, 196140 Russia

Key words: color-magnitude diagram, Galactic solar neighborhoods, giant and subgiant stars,
proper motions and radial velocities.

Based on multicolor photometry from the 2MASS and Tycho-2 catalogues, we have produced
a sample of 38 368 branch red giants that has less than 1\% of admixtures and is
complete within 500 pc of the Sun. The sample includes 30 671 K giants, 7544 M giants,
49 C giants, and 104 suspected supergiants or S stars.
The photometric distances have been calculated for K, M, and C stars with an accuracy
of 40\%. Tycho-2 proper motions and PCRV radial velocities are used to analyze the
stellar kinematics.
The decrease in the stellar distribution density with distance from the Galactic
equator approximated by the barometric law, contrary to the Besancon model of the
Galaxy, and the kinematic parameters calculated using the Ogorodnikov--Milne model
characterize the overwhelming majority of the selected K and M giants as disk stars
with ages of more than 3 Gyr.
A small number of K and M giants are extremely young or, conversely, thick-disk ones.
The latter show a nonuniform distribution in the phase space of
coordinates and velocities, arguing against isothermality and full relaxation of
the disk and for the theory of dynamical streams or superclusters.
The spatial distribution and kinematics of the selected C stars force us
to consider them as asymptotic branch giants with masses of more than
2 $M_{\odot}$ and ages of less than 2 Gyr probably associated with the Gould Belt.
The offset of the Sun above the Galactic equator has been found from the distribution
of stars to be $13\pm2$ pc, which coincides with the previously obtained value for the
clump red giants.

\newpage
\section*{INTRODUCTION}

In this paper, we continue to investigate the spatial distribution and kinematics of stars of
various types within the kiloparsec of the Galaxy nearest to the Sun.
Previously, based on multiband photometry from the Tycho-2 (H\o g et al. 2000) and 2MASS (Skrutskie
et al. 2006) catalogues, Tycho-2 proper motions, and PCRV radial velocities (Gontcharov 2006) and
using calibrations based on data from the Hipparcos catalogue, the first (ESA 1997) and new (van
Leeuwen 2007) versions, we produced and analyzed samples of OB stars (Gontcharov 2008a), red giant
clump (RGC) stars (Gontcharov 2008b), and subdwarfs (Gontcharov 2010). These works, along
with the corresponding Monte Carlo simulations (Gontcharov 2009a), showed that stars of some
types could be successfully selected from modern photometric and astrometric surveys without any
spectral classification for their further studies.

Here, we selected and investigated stars belonging to the red giant branch (RGB) in its reddest
part, i.e., giants with effective temperatures lower than those for RGC ones.
According to the spectral classification, the selected stars are of types later than K3I--K3III.
However, the spectral classification is known only for some of the selected stars and was not used
here to determine any quantities.

\section*{THE SELECTION OF STARS}

\subsection*{Input Photometry}

The accuracy of the photometry in Tycho-2 (the $B_T$ and $V_T$ bands) and 2MASS
(the infrared (IR) $J$, $H$, and $Ks$ bands) deteriorates for faint stars. For
the selection of RGB stars, we used 718 000 stars with $B_{T}<11.7^{m}$ and $V_{T}<11^{m}$.
The median accuracy of their photometry is $\sigma(B_{T})=0.04^{m}$,
$\sigma(V_{T})=0.03^{m}$. We imposed no constraints on the $J$, $H$, and
$Ks$ magnitudes, because here we consider red stars that are rather bright in the infrared.
At $Ks>5.5^{m}$, the median accuracy is
$\sigma(J)=0.02^{m}$, $\sigma(H)=0.03^{m}$, $\sigma(Ks)=0.02^{m}$.

However, the near-IR photometry for the brightest stars ($Ks<5.5^{m}$) is inaccurate, but there is
accurate far-IR photometry for them from the IRAS catalog (IRAS 1988). We cross-identified Tycho-2
and IRAS using the SIMBAD database in Strasbourg. For red giants with accurate photometry,
the $Ks$ magnitude was calibrated as a function of the logarithm with base 2.512 of the IR flux
at wavelength 12 microns (below referred to as $F12$) taken from IRAS:
$Ks=0.0087(\log_{2.512}(F12))^3)-0.0371(\log_{2.512}(F12))^2)-0.95(\log_{2.512}(F12)))+4.2$.
The calibration gave $\sigma(Ks)=0.18^m$. Owing to this calibration, all bright stars were retained
in the sample.

In the sample, we retained only single stars and bright components of multiple systems with
$\sigma(\mu_{\alpha}\cos\delta)<5$, $\sigma(\mu_{\delta})<5$ mas yr$^{-1}$, i.e., 97\% of
the sky stars brighter than $B_{T}=11.7^{m}$ and $V_{T}=11^{m}$.
The median accuracy is $\sigma(\mu_{\alpha}\cos\delta)=\sigma(\mu_{\delta})=1.3$ mas yr$^{-1}$.

\subsection*{The $(V_T-Ks)$ -- $(B_T-V_T)$ Diagram}

We selected and classified RGB stars using two-color diagrams, primarily the
$(V_{T}-Ks)$ -- $(B_{T}-V_{T})$ diagrams with the most accurate color indices. Other
two-color diagrams and diagrams with three-color indices Q were used for checking.

Part of the $(V_{T}-Ks)$ -- $(B_{T}-V_{T})$ diagram with the RGC and the RGB is shown in Fig. 1.
To present the theoretical data on this diagram, we recalculated the Johnson $B_j$ , $V_j$,
and $K_j$ magnitudes to the $B_T$, $V_T$, and $Ks$ magnitudes by taking into account the
relations $(B_{T}-V_{T})=(B_j-V_j)/0.78$ for giant stars (Gontcharov 2008b) and
$(V_{T}-Ks)=(V_j-K_j)+0.09(B_{T}-V_{T})$ (ESA 1997). Both relations introduce
an error of no more than 0.05$^{m}$.

The main sequence and the giant branch on this diagram are not separated.
However, as we showed previously (Gontcharov 2009a), there are actually no
dwarfs among the reddest Tycho-2 stars, because Tycho-2 is limited in apparent
magnitude (below, we discuss the elimination of the remaining dwarfs
as well). To produce a pure sample of RGB stars free not only from dwarfs but also
from RGC stars, we imposed the constraints $(J-Ks)>0.8-\sigma(J-Ks)$,
$(B_{T}-V_{T})>1.2^{m}$, $(V_{T}-Ks)>2.8$ and $(B_{T}-V_{T})>1.86-0.086(V_{T}-Ks)$
on the sample (the last two conditions are indicated by the dashed straight
lines in the left bottom part of the figure) that follow from the theoretical evolutionary tracks
by Girardi et al. (2000) for stars older than 1 Gyr (the younger RGC stars of the color under
consideration are extremely rare).

These constraints do not eliminate noticeably reddened RGC stars. The following approach was
applied for them. The reddening line of length $E_{(B_T-V_T)}=0.5^{m}$ is indicated by the vector
in the left part of the figure according to Gontcharov (2008a). The ``cloud'' of points in the
figure is smeared by the reddening in this direction. Assuming the reddening of stars at high
latitudes to be small ($E_{(B_T-V_T)}<0.3^{m}$) and selecting such stars, we can calculate
the average dependence of $(B_{T}-V_{T})$ on $(V_{T}-Ks)$ for them. It is not shown in the
figure, because it coincides with the theoretical and mean empirical colors obtained by
Pickles (1998) based on the spectral energy distribution for unreddened solar-metallicity
K3III, K4III, K5III, M0III, M1III, M2III, M3III, M4III, M5III, M6III, M7III, M8III,
M9III, M10III stars that are indicated by the large open squares in the figure, respectively,
from left to right. Shifting this dependence along the reddening line, we obtain the line of
normal colors for M stars indicated in the figure by the thin solid black curve below the main
cloud of points: $(B_{T}-V_{T})=0.0183(V_{T}-Ks)^3-0.3594(V_{T}-Ks)^2+2.15(V_{T}-Ks)-2.3$.
Taking into account the accuracy of the photometry, we obtain the line below which the
stars cannot belong to the RGB and were rejected as reddened RGC stars:
$(B_{T}-V_{T})=0.0183(V_{T}-Ks)^3-0.3484(V_{T}-Ks)^2+2(V_{T}-Ks)-2$. This
polynomial is indicated by the dashed line at the bottom of the figure. The selected RGB stars are
indicated in the figure by the black symbols and the rejected extraneous stars, mostly the RGC ones,
are indicated by the gray points below and leftward of the three dashed lines.

An important feature of this diagram is the kink in the dependence of $(B_{T}-V_{T})$ on
$(V_{T}-Ks)$ for RGB stars seen in the figure that separates the K
and M stars. This nonlinearity of the dependence is explained by a difference in spectral
energy distribution for K and M stars, namely the presence of strong
absorption lines in the spectra of M stars in the $V_T$ band.

According to Pickles (1998), the nonsolar metallicity stars and bright giants at the level of
accuracy under consideration are lost among the solar metallicity giants and cannot be
distinguished on this diagram. In contrast, according to the Tycho Spectral Types (TST) catalogue
(Wright et al. 2003), the supergiants and S stars (inseparable photometrically and below designated
together as S/supergiants) and carbon stars (below referred to as C ones) must have slightly
larger $(B_T-V_T)$ color indices at the same $(V_T-Ks)$. Although an admixture of
highly reddened or peculiar KIII and MIII stars is possible here, we, nevertheless, distinguished
two groups of stars: (a) more likely C stars (marked by the diamonds at the top of the figure)
and (b) S/supergiants (marked by the filled squares). Taking into account the spectral energy
distribution, we assume that a star is more likely a C one than an S/supergiant one when all of
the following conditions are met: $(B_{T}-V_{T})>0.27(V_{T}-Ks)+1.14$
(indicated by the thick gray straight line at the top
of the figure), $(J-Ks)>1.35^{m}$ and $(V_{T}-J)<7(J-H)-1.9$. The suspected S/supergiants and
C stars are separated from the KIII and MIII stars
by the polynomial $(B_{T}-V_{T})=0.009(V_{T}-Ks)^3-0.1828(V_{T}-Ks)^2+1.1552(V_{T}-Ks)+0.2$.
The latter is indicated in the figure by the thick gray curve and actually reflects a reasonable
assumption about the interstellar extinction in the region of space under
consideration: $A_V<3^{m}$. The KIII and MIII stars were separated along the reddening line
$(B_{T}-V_{T})=0.365(V_{T}-Ks)+0.25$ (the gray straight line at the center of the figure),
although this separation is arbitrary, because the properties of stars change
gradually in the sequence of K--M types.

\subsection*{The Elimination of Dwarfs}

To eliminate dwarfs, we used reduced proper motions:
$M'_{VT}=V_{T}+5+5\cdot~lg(\mu)$ and the analogous
$M'_{Ks}$, where $\mu=(\mu_{\alpha}^2\cos\delta+\mu_{\delta}^2)^{1/2}$ is the proper motion
in arcseconds. The constraints $M'_{VT}<8.7$ and $M'_{Ks}<5$ were imposed on the sample.
In this way, we eliminated 44 suspected dwarfs. As expected, the number of dwarfs in the range of
colors under consideration is very small. Thus, RGB stars were actually selected using only the
colors, as distinct from the selection of RGC stars in Gontcharov (2008b), where
the reduced proper motions played a key role.

\section*{CHARACTERISTICS OF STARS}

\subsection*{Reddening and Extinction}

The reddening $E_{(B_T-V_T)}$ and extinction $A_V$ (the extinctions in the $V_j$ and $V_T$ bands
are equal with a sufficient accuracy) for the stars under consideration were used to calibrate the
absolute magnitudes $M_{VT}$ from the $(V_{T}-Ks)$ color and to calculate the photometric
distances $R_{ph}$ from the formula
\begin{equation}
\label{rph}
\log(R_{ph})=(V_{T}-M_{VT}+5-A_V)/5.
\end{equation}
As we show below, in both cases, the accuracy is determined by the natural scatter
$\sigma(M_{VT})=0.8^{m}$, not by a much more accurate correction for the extinction.

The extinction for each selected star was estimated by three different methods:
\begin{itemize}
\item From the formula $A_{EBV}=3.1E_{(B_T-V_T)}$, where the reddening $E_{(B_{T}-V_{T})}$ was
calculated as the shift of the star from the line of normal colors mentioned above. Given the
accuracy of the photometry and variations in the extinction coefficient of about 3.1, the accuracy
of this extinction estimate is $\sigma(A_{EBV})\approx0.3^{m}$.
\item From the deviation of the spectral energy distribution for the star from the unreddened one,
i.e., from a linear combination of photometric magnitudes. Previous (Gontcharov 2008b), we showed
that the extinction $A_{BVJK}=0.96(-1.89B_{T}+2.89V_{T}+0.25J-1.25Ks-0.26)$.
For stars without strong lines in their spectra, the accuracy of this estimate is
determined by the accuracy of the photometry and is also $\sigma(A_{BVJK})\approx0.3^{m}$.
\item As a function of the Galactic coordinates, $A_{G}=f(l, b, R_{ph})$, from our three-dimensional
extinction model (Gontcharov 2009b). In this case, $A_{G}$ and $R_{ph}$ are refined by iterations:
$M_{VT}$ is calibrated from $(V_{T}-Ks)$, as is shown below; subsequently, we calculate
$\log(R_{ph})=(V_{T}-M_{VT}+5)/5$ in the first approximation, then calculate $A_{G}=f(l, b, R_{ph})$,
and recalculate $R_{ph}$ in the second approximation using Eq. (1). The iterations converge rapidly,
because $A_V$ in Eq. (1) is small compared to the remaining quantities. Within 500 pc of the Sun
(i.e., for most of the stars under consideration), the accuracy of the model
$\sigma(A_{G})\approx0.3^{m}$ was estimated previously (Gontcharov 2009b) by comparing it with
three catalogs of individual extinctions for stars. We also showed in the same paper that the
accuracy of correction for the extinction 0.3$^{m}$ typical of all methods is
a difficult-to-surmount limit due to the corresponding natural scatter of individual extinctions
for neighboring stars.
\end{itemize}

Comparison of $A_{EBV}$, $A_{BVJK}$ and $A_{G}$ for the selected stars showed the following:
\begin{itemize}
\item $A_{EBV}$ is implausibly large for C stars and S/supergiants, because the relation between
$(B_{T}-V_{T})$ and $(V_{T}-Ks)$ is different from that for K and M stars;
\item because of strong spectral lines, $A_{BVJK}$ is implausibly large for many of the M stars
and S/supergiants and is negative for many of the C stars;
\item $A_{G}$ has been overestimated for stars farther than 500 pc, outside the Gould Belt:
for example,
the mean differences for K stars $\overline{A_{G}-A_{BVJK}}=\overline{A_{G}-A_{EBV}}=0.3^{m}$;
\item for the remaining stars, different extinction estimates agree at the level of the mentioned
limit 0.3$^{m}$: the standard deviation is $\sigma(A_{G}-A_{BVJK})=0.3^{m}$ for
K, C stars and S/supergiants within 500 pc of the Sun, $\sigma(A_{G}-A_{EBV})=0.3^{m}$
for K and M stars within the same radius, and $\sigma(A_{BVJK}-A_{EBV})=0.3^{m}$ for
K stars in the entire region of space.
\end{itemize}

Below, we use $A_{EBV}$ for K and M stars and $A_{G}$ for C stars. No extinction was used for
S/supergiants, because $M_{VT}$ and $R_{ph}$ were not calculated due to
the scarcity of data for the calibration of $M_{VT}$ from $(V_T-Ks)$.

\subsection*{Absolute Magnitudes and Distances}

There are 38 368 RGB stars in the final sample. The new Hipparcos reduction (van Leeuwen 2007)
contains 12 208 of the selected stars, but only 2900 of them have parallaxes with a relative
accuracy better than 20\%: 536 M stars, 2354 K stars, 9 C stars, and 1 S star.
We used them to calculate the absolute magnitude $M_{VT}$, which was then calibrated from the
$(V_{T}-Ks)$ color. In this case, $M_{VT}$ and $(V_{T}-Ks)$ were corrected for the previously
found extinction and reddening with $E_{(V_{T}-Ks)}=2.74E_{(B_T-V_T)}$
from Fitzpatrick and Massa (2007), who also estimated the relative accuracy of this relation
to be better than 2\%.

To estimate the effects of the Malmquist and Lutz--Kelker biases (Perryman 2009, pp. 209--211) on the
dependence of $M_{VT}$ on distance and $(V_{T}-Ks)$, we performed Monte Carlo simulations.
We used Gaussian and uniform distributions implemented by the Microsoft Excel 2007 random number
generator, whose general description was given by Wichman and Hill (1982). We generated 65 000
model stars for which we specified the following:
\begin{itemize}
\item a uniform distribution in rectangular Galactic coordinates $X$ and $Y$ within 1100 pc of
the Sun and a Gaussian distribution in $Z$ with a standard deviation of 200 pc (this corresponds
with a sufficient accuracy to the distribution of comparatively old disk stars according
to the Besancon model of the Galaxy (BMG) (Robin et al. 2003));
\item a uniform distribution in the range $3^{m}<(V_{T}-Ks)<8^{m}$.
\end{itemize}
Subsequently, we calculated the following:
\begin{itemize}
\item $M_{VT}$ with a mean $-0.11(V_{T}-Ks)^3+2(V_{T}-Ks)^2-11.4(V_{T}-Ks)+20.4$ and a standard
deviation of 0.8$^{m}$ in accordance with the theoretical isochrones for a mixture of stars with
ages of 10$^9$--10$^{10}$ yr (Girardi et al. 2000) with allowance
made for $V_{T}\approx~V_j+0.2^{m}$;
\item the true distance $R=(X^2+Y^2+Z^2)^{1/2}$, the
Galactic coordinates $\tan(l)=Y/X$, $\tan(b)=Z/(X^2+Y^2)^{1/2}$,
\item the interstellar extinction $A_V$ from our analytical extinction model (Gontcharov 2009b) and
the reddening $E_{(V_T-Ks)}=A_V/3.1\cdot2.74$;
\item $V_{T}=M_{VT}-5+5\lg(R)+A_V$ and the photometric error $\sigma(V_{T})=0.01e^{0.3V_{T}}$
in accordance with the Tycho-2 characteristics;
\item the true parallax $\pi=1000/R$ (where R in pc and $\pi$ in mas)
and its measured value with a standard
deviation of 1 mas relative to the true value.
\end{itemize}
In accordance with the limitation of the Hipparcos catalogue (ESA 1997, p. 131), we retain only
the stars with $V_{T}<7.3+1.1|\sin(b)|$ and only the stars with a relative accuracy of the parallax
better than 20\%.

Our simulations showed both a considerable decrease in the mean value of $M_{VT}$ and some change in
the dependence of $M_{VT}$ on $(V_{T}-Ks)$ with distance. However, for stars with $\pi>3.3$
mas these effects do not exceed 0.1$^{m}$. This conclusion was confirmed by
our analysis of the change in the dependence of $M_{VT}$ on $(V_{T}-Ks)$ with distance for the real selected K and
M stars.

As a result, we calibrated $M_{VT}$ as a function of $(V_{T}-Ks)$ for 2167 K and M stars with a relative
accuracy of the parallax better than 20\% and $\pi>3.3$ mas: the calibration curve
$M_{VT}=0.124(V_{T}-Ks)^2-1.16(V_{T}-Ks)+2.24$ found is shown in Fig. 2 as the thick black line.
For C stars, we adopted the same dependence but brighter by 1$^{m}$. In Fig. 2, the K and M stars used for the
calibration are marked by the crosses and snowflakes, respectively. The C stars (circles) and one S star (square)
with a relative accuracy of the parallax better than 20\% that were not used for the calibration are also marked here.

For comparison with the observational data, Fig. 2 shows the theoretical isochrones for solar-metallicity
stars from Girardi et al. (2000): the dashed and solid thin curves indicate the RGB and the asymptotic
giant branch, respectively; the upper dashed and solid curves correspond to an age of 10$^9$ yr and the lower
curves correspond to 10$^{10}$ yr. Stellar masses of about 2.3 $M_{\odot}$ for $10^9$ yr and about
1.1 $M_{\odot}$ for $10^{10}$ yr correspond to the segments of the isochrones falling into
the region filled with the selected stars shown in the figure. The massive young stars of the sample are
apparently on the asymptotic giant branch, while the low-mass old ones are on the RGB. The theoretical
isochrones do not quite correspond to the positions of the reddest stars from the sample, which is apparently
indicative of shortcomings of the theory. On the whole, the positions of the selected stars relative
to the isochrones suggest that disk stars with ages from 1 to 10 Gyr dominate in the sample. The picture
is complicated by the fact that metallicity variations displace significantly the stars and isochrones. As a
result, it is difficult to draw conclusions about the age and metallicity of a specific star from its position on
the diagram.

For K, M, and C stars, the standard deviation of $M_{VT}$ from the calibration curve is 0.8$^{m}$. This allows
us to calculate the photometric distances $R_{ph}$ from Eq. (1) with a relative accuracy of 40\% and the corresponding
rectangular Galactic coordinates $X$, $Y$, $Z$.

\subsection*{Comparison with Known Classification}

Out of the 38 368 selected stars, 31 049 (81\%) have spectral classification either from the Hipparcos
Input Catalogue (HIC) (Turon et al. 1993) or from the TST. For many stars, the classification from
these catalogues differs or is designated as dubious. Among the stars selected as S/supergiants, five were
previously classified as type C. This classification can apparently be trusted and below we consider them to
be C stars. As a result, we attributed 30 671, 7544, 104, and 49 stars to types K, M, S/supergiants, and
C, respectively.

The main statistics of the spectral classification is the following: 6029 of the 7544 stars selected as
type M have the old classification, 5200 (86\%) of them were classified as M at least in one catalogue,
and most of the remaining stars were classified as K; 24 872 of the 30 671 stars selected as type K have the
old classification, 21 589 (87\%) of them were classified as K at least in one catalogue, and most of the remaining
stars were classified as M; for S/supergiant, these numbers are, respectively, 104, 99, and 45
(45\%) (for many of the remaining stars, the luminosity class is unknown); for type C, the numbers are 49, 49,
46 (94\%). The correctness of the classification for the selected S/supergiants is confirmed by their strong
concentration to the Galactic plane: $|b|<20^{\circ}$ for all stars. Thus, since the derived classification for the
overwhelming majority of the selected stars coincides with the old one, the applied method of classification
based on broadband multicolor photometry should be recognized as successful.

Moreover, for some stars, the new classification based on photometry is more accurate than the old
spectral one. For example, 191 and 135 selected stars were previously classified as dwarfs and subgiants,
respectively. This classification cannot be trusted, because the Hipparcos parallaxes for 56 dwarfs and
30 subgiants are known with a relative accuracy better than 50\%. This allows their luminosity class to
be checked by the absolute magnitude: on average, $\overline{M_{VT}}=-0.1\pm1.2$ for dwarfs and
$\overline{M_{VT}}=0.4\pm1.2$ for subgiants. These values correspond to those for
giants and are totally unsuitable for dwarfs ($\overline{M_{VT}}\approx8$) and subgiants
($\overline{M_{VT}}\approx3$) according to Hipparcos data.

\subsection*{The Spatial Distribution of Stars}

The distribution of the selected stars in projection onto the $XY$, $XZ$, and $YZ$ planes is shown in Fig. 3:
(a) type K, (b) type M, and (c) type C. The radius of the region where the sample is complete can be
estimated from the change in the stellar distribution density and from the maximum of the mean interstellar
extinction that is reached at the edge of this region: further out, the mean extinction decreases, because
the sample becomes incomplete mainly through the loss of stars with large extinction. Both approaches
give the same estimate: for all types of stars, the sample is complete up to 500 pc. In this space,
the distribution density of K and M stars decreases toward the Galactic anticenter as for typical disk stars
in accordance with the BMG. In contrast, C stars dominate in the second and third Galactic quadrants
and are encountered here at greater distances than in the first and fourth quadrants. The Gould Belt stars
are distributed precisely in this way (Perryman 2009, pp. 324--328). For this reason, below we discuss the
evolutionary status and ages of C stars.

``Flaws'' are noticeable in the distribution of stars of all types on the $XZ$ plots because of the interstellar
extinction in the Gould Belt, which is oriented roughly in this plane with an inclination of 17$^{\circ}$ to the
equator (Gontcharov 2009b).

The decrease in the stellar distribution density with distance from the equator can theoretically be
proportional to $sech^{2}(2Z/Z_{0})$ (Girardi et al. 2005) or correspond to the complex dependence from the
BMG or the barometric law $D_{0}\cdot~e^{-|Z|/Z_{0}}$, where $D_{0}$ is the density in the equatorial plane
and $Z_{0}$ is the distance from this plane at which the density decreases by a factor of $e$ or the half-thickness
of a homogeneous layer of stars (Parenago 1954, p. 264). Figure 4 shows the distribution of the selected (a) K
and (b) M stars as a function of the $Z$ coordinate in the vertical cylinder elongated along the $Z$ axis with
a radius of 400 pc around the Sun and the curves of the bolometric law fitting best this distribution:
$D_{0}=1262$, $Z_{0}=263$ pc for type K and $D_{0}=555$, $Z_{0}=217$ pc for type M. The values of $Z_{0}$ found are
typical of rather old Galactic disk stars according to the BMG. The lower value for M stars most likely
suggests not their relative youth but the possible contamination of their sample by young forming stars, C
stars, and supergiants. This fit closely corresponds to the observed distribution at all $Z$, except for the region
$-100<Z<100$ pc, where less than 100 stars of each type produce an excess that can also be fitted by
the barometric law with $Z_{0}=50$ pc. These stars are apparently an admixture of young stars. This excess
near the equator cannot be removed even by reducing the $Z$ range used for fitting. It is also important that
fitting the data by the square of the hyperbolic secant or by the dependence from the BMG for any reasonable
characteristics of the thin and thick disks does not give a better result than the barometric law. Thus,
because of their comparatively small number, the thick-disk giants do not manifest themselves clearly
among the selected stars. However, having extreme velocities, as we show below, they manifest themselves
when averaging the kinematic characteristics of the stars in the same spatial cells.

The discrepancy between the $Z$ distribution found and the dependence from the BMG apparently suggests
the absence of isothermality and full relaxation of the Galactic disk assumed in the BMG. This is
an expected result in connection with the detection of several kinematically isolated groups among comparatively
old red giants that are superclusters or dynamical streams by Famaey et al. (2005).

The asymmetric distribution of stars in the cylinder under consideration southward and northward of
the Galactic equator allows the offset of the Sun to be calculated: 13$\pm2$ pc; this corresponds to the value
found previously (Gontcharov 2008b) from RGC stars.

\subsection*{Stellar Kinematics}

The present views of the Galactic evolution suggest the relationship between the age of stars, their
metallicity, the distance from the Galactic equator, the velocity dispersion, and the velocity component $V$.

The proper motions and radial velocities allow one to calculate the velocities of stars in Galactic $U$, $V$, $W$
coordinates and to analyze their kinematics by comparing the results with the analysis of a sample of red
giants from Hipparcos (Famaey 2005) and a sample of RGC stars from Tycho-2 (Bobylev et al. 2009).

The radial velocities $V_r$ from the PCRV catalog are known with an accuracy better than 5 km s$^{-1}$
for the 3653 selected stars: 2737 K, 852 M, 41 C, and 23 S/supergiant stars (since the distances for
the S/supergiants were not calculated, they are not considered below). The absence of observational selection
in the distribution of these stars on the celestial sphere frees the kinematic characteristics being
determined from biases, as distinct from the sample by Famaey et al. (2005) containing only Northern-Hemisphere stars.

The accuracy of the calculated $U$, $V$, $W$ is actually determined by the accuracy of the derived distances.
It is about 15 km s$^{-1}$ for each velocity component, when using both trigonometric and photometric distances.
This also determines the accuracy of the kinematic parameters obtained below.

Consider the dependences of the mean velocity components and velocity dispersions on $Z$ coordinate
in the vertical cylinder elongated along the $Z$ axis with a radius of 500 pc around the Sun for (a) type K and
(b) type M shown in Fig. 5. The accuracy is indicated by the vertical bars. The results from photometric
distances and Hipparcos parallaxes with a relative accuracy of the parallax better than 20\% are indicated
by the solid and dashed lines, respectively. We see that everywhere, except the extreme $Z$, the results
from the two types of distances agree well between themselves. The disagreements at the extreme $Z$
are explained by the influence of few thick-disk (and, possibly, halo) stars with extreme velocities located at
the boundary of the region of space under consideration and included or not included in the sample depending
on whether the trigonometric or photometric distances are used.

The same stars causes a \emph{nonuniform} decrease in the velocity component $V$ and an increase in the
velocity dispersion with distance from the equator, for example, at $-450<Z<-350$ pc for K stars and
$Z\approx-450$, $-250$, and $-50$ pc for M stars. The few (among the selected stars) thick-disk (and, possibly,
halo) giants, which, as has been mentioned above, are distributed in the phase space of coordinates and
velocities as isolated groups, dynamical streams or superclusters (Famaey et al. 2005), manifest themselves
in this way. However, disk giants younger than 10 Gyr dominate in the entire region of space under
consideration: the velocities and velocity dispersions averaged over the spatial cells are almost everywhere
within the limits established by the BMG for thin-disk stars
($\sigma(U)<50$, $\sigma(V)<35$, $\sigma(W)<25$, $V>-35$ km s$^{-1}$). In subsequent studies, we will consider
specific stars and groups of stars from the thick disk that cause oscillations of the mean velocities and
velocity dispersions.

\subsection*{The Ogorodnikov--Milne Model}

The kinematic parameters of the selected stars were determined within the framework of a linear
Ogorodnikov--Milne model described in detail by Bobylev et al. (2009). The solar velocity components
$U_\odot$, $V_\odot$, $W_\odot$ relative to the centroid of the stars
under consideration and nine partial derivatives of the velocity with respect to the distance forming the
displacement matrix $M$ were determined from the known (trigonometric or photometric) distances $R$,
celestial coordinates, radial velocities, and proper motions of the selected stars. In this case, the
conditional equations solved by the least-squares method are

$$
\displaylines{\hfill
 V_r=-U_\odot\cos~b\cos~l-V_\odot\cos~b\sin~l\hfill\llap{(2)} \cr\hfill
-W_\odot\sin~b+R (\cos^2 b\cos^2 l M_{ux}
\hfill\cr\hfill
 +\cos^2 b\cos l\sin l M_{uy}+\cos b\sin b\cos l M_{uz}
\hfill\cr\hfill
 +\cos^2 b\sin l\cos l M_{vx}+\cos^2 b\sin^2 l M_{vy}
\hfill\cr\hfill
 +\cos b\sin b\sin l M_{vz}+\sin b\cos b\cos l M_{wx}
\hfill\cr\hfill
 +\cos b\sin b\sin l M_{wy}+\sin^2 b M_{wz}),
\hfill\cr\hfill
 4.74 R \mu_l\cos b= U_\odot\sin l-V_\odot\cos l+ ~~~\hfill\llap{(3)} \cr\hfill
 +R (-\cos b\cos l\sin l  M_{ux} -\cos b\sin^2 l  M_{uy}-
\hfill\cr\hfill
 -\sin b \sin l  M_{uz} +\cos b\cos^2 l M_{vx}+
\hfill\cr\hfill
 +\cos b\sin l\cos l  M_{vy}+\sin b\cos l  M_{vz}),
\hfill\cr\hfill
4.74 R \mu_b= U_\odot\cos l\sin b+V_\odot\sin l\sin b
\hfill\llap{(4)} \cr\hfill
-W_\odot\cos b+R (-\sin b\cos b\cos^2 l M_{ux}-
\hfill\cr\hfill
 -\sin b\cos b\sin l \cos l M_{uy}-\sin^2 b \cos l  M_{uz}
\hfill\cr\hfill
 -\sin b\cos b\sin l\cos l M_{vx}-\sin b\cos b\sin^2 l  M_{vy}
\hfill\cr\hfill
 -\sin^2 b\sin l  M_{vz}+\cos^2 b\cos l M_{wx}
\hfill\cr\hfill
 +\cos^2 b\sin l M_{wy}+\sin b\cos b  M_{wz}).
\hfill }
$$

We then calculate the Oort constants
 $A=0.5(M_{uy}+M_{vx})$,
 $B=0.5(M_{vx}-M_{uy})$,
 $C=0.5(M_{ux}-M_{vy})$,
 $K=0.5(M_{ux}+M_{vy})$,
the vertex deviation $l_{xy}$: $\tan 2 l_{xy}=-C/A$ and the angular
velocity of Galactic rotation $\Omega_{R_0}=B-A=-M_{uy}$.

The results of the solution, along with the number
of stars, the mean distances $\overline{R}$, and the velocity
dispersions $\sigma(U)$, $\sigma(V)$, $\sigma(W)$, are presented in the
table for types K, M, and C in two variants: with the
trigonometric, $R_{hip}$, and photometric, $R_{ph}$, distances.

We see that the results for $R_{hip}$ and $R_{ph}$ agree, within the error limits. The Oort constant $A$ and
$B$ agree, within the error limits, with the universally accepted ones, for example, those found by Bobylev
et al. (2009) for the RGC, $A=15.9\pm0.2$, $B=-12.0\pm0.2$. The vertex deviation and the constant $C$
slightly differ from $l_{xy}=7.0\pm0.3$, $C=-3.9\pm0.2$ found by Bobylev et al. The constant $K$ shows
compression for the set of K stars and expansion for type M. When an admixture of extremely young
stars involved in the expansion of the Gould Belt is present among the M stars, this result is explainable
and agrees with the result by Bobylev (2004) and the review by Perryman (2009, pp. 324--328).

Comparison of the velocity dispersions and $V_{\odot}$ with the BMG leads us to conclude that the K and
M stars have similar kinematics typical of disk stars with ages of more than 3 Gyr.

The kinematics of the C stars is completely different and is typical of disk stars with ages of less than
2 Gyr. Taking into account the spatial distribution of these stars noted above, which may be associated
with the Gould Belt, and their positions relative to the isochrones on the $(V_{T}-Ks)$ -- $M_{VT}$
diagram (Fig. 2), we can conclude that these are relatively massive ($>2M_{\odot}$) asymptotic branch giants.

Perryman (2009, pp. 449--452) gave an overview of the present ideas about the C stars including two
main groups with a distinctly different evolutionary status. It is hypothesized that the hotter R-subtype
carbon stars belong to comparatively old clump red giants with an approximately solar mass, while the
cooler N-subtype carbon stars belong to younger asymptotic branch giants with masses of $1.5-4 M_{\odot}$.
Knapp et al. (2001) showed that many of the N stars had so far been erroneously classified as R, although
these subtypes must occupy different regions of the color--absolute magnitude diagram. The characteristics
of the set of C stars that we obtained here (the uniqueness of the stellar positions on two-color
diagrams, the accuracy of the velocities, and the completeness of the sample of 49 selected stars, 41 with
velocities, and 46 with spectral classification) confirm that the C type in the range of colors under consideration
actually includes only comparatively young and massive asymptotic branch giants that apparently belong
to subtype N without exception and the applied method is efficient for their identification.

%¹¹¹¹¹¹¹¹¹¹¹¹¹¹¹¹¹¹¹¹¹¹¹¹¹¹¹¹¹¹¹¹¹¹¹¹¹¹¹¹¹¹¹¹¹¹¹¹¹¹¹¹¹¹¹¹¹¹¹¹¹¹¹¹¹  %% Òàáëèöà 1
\begin{table*}[!h]
\def\baselinestretch{1}\normalsize\footnotesize
\caption[]{Kinematic parameters of the selected stars
}
\label{kine}
\[
\begin{tabular}{l|r|r|r|r|r|r}
\hline
\noalign{\smallskip}
   & \multicolumn{3}{c|}{$R_{hip}$} &  \multicolumn{3}{c|}{$R_{ph}$}  \\
\hline
Parameter & K & M & C & K & M & C \\
\hline
\noalign{\smallskip}
Number of stars                   & 2360          & 685         & 29           & 2737          & 852         & 41  \\
$\overline{R}$, pc                & 326           & 329         & 459          & 261           & 298         & 273 \\
$\sigma(U)$, km s$^{-1}$          & $37\pm$0.6    & 40$\pm$0.9  & 28$\pm$4     & 34$\pm$0.5    & 37$\pm$0.7  & 21$\pm$3  \\
$\sigma(V)$, km s$^{-1}$          & $25\pm$0.6    & 30$\pm$0.9  & 18$\pm$4     & 27$\pm$0.5    & 30$\pm$0.7  & 14$\pm$3 \\
$\sigma(W)$, km s$^{-1}$          & $20\pm$0.6    & 22$\pm$0.9  & 10$\pm$4     & 20$\pm$0.5    & 21$\pm$0.7  & 6$\pm$3   \\
U$_{\odot}$, km s$^{-1}$          & $6\pm$0.6     & 9$\pm$0.9   & 13$\pm$4     & 6$\pm$0.5     & 9$\pm$0.7   & 12$\pm$3 \\
V$_{\odot}$, km s$^{-1}$          & $20\pm$0.6    & 24$\pm$0.9  & 8$\pm$4      & 19$\pm$0.5    & 23$\pm$0.7  & 8$\pm$3  \\
W$_{\odot}$, km s$^{-1}$          & $7\pm$0.6     & 8$\pm$0.9   & 6$\pm$4      & 6$\pm$0.5     & 7$\pm$0.7   & 4$\pm$3  \\
M$_{ux}$, km s$^{-1}$ kpc$^{-1}$  & $-3\pm$4      & $-1\pm$6    & $-4\pm$30    & $-4\pm$4      &  $5\pm$6    & $32\pm$20 \\
M$_{uy}$, km s$^{-1}$ kpc$^{-1}$  & $29\pm$4      & $27\pm$6    & $19\pm$30    & $33\pm$4      & $23\pm$6    & $30\pm$20 \\
M$_{uz}$, km s$^{-1}$ kpc$^{-1}$  & $-8\pm$4      & $10\pm$6    & $15\pm$30    & $-14\pm$4     & $6\pm$6     & 31$\pm$20 \\
M$_{vx}$, km s$^{-1}$ kpc$^{-1}$  & $3\pm$4       & $1\pm$6     & $-10\pm$30   & $0\pm$4       & $-2\pm$6    & 12$\pm$20  \\
M$_{vy}$, km s$^{-1}$ kpc$^{-1}$  & $1\pm$4       & $10\pm$6    & $-11\pm$30   & $-8\pm$4      & $4\pm$6     & $-11\pm$20 \\
M$_{vz}$, km s$^{-1}$ kpc$^{-1}$  & $-5\pm$4      & $-5\pm$6    & $19\pm$30    & $-7\pm$4      & $-8\pm$6    & $19\pm$20 \\
M$_{wx}$, km s$^{-1}$ kpc$^{-1}$  & $1\pm$4       & $-1\pm$6    & $-2\pm$30    & $2\pm$4       & $7\pm$6     & $-2\pm$20 \\
M$_{wy}$, km s$^{-1}$ kpc$^{-1}$  & $-1\pm$4      & $-2\pm$6    & $-8\pm$30    & $-5\pm$4      & $0\pm$6     & $0\pm$20  \\
M$_{wz}$, km s$^{-1}$ kpc$^{-1}$  & $1\pm$4       & $2\pm$6     & $24\pm$30    & $-3\pm$4      & $7\pm$6     & $17\pm$20   \\
A, km s$^{-1}$ kpc$^{-1}$         & $16\pm$2      & $14\pm$4    & $4.5\pm$20   & $16.5\pm$2    & $10.5\pm$4  & $21\pm$15 \\
B, km s$^{-1}$ kpc$^{-1}$         & $-13\pm$2     & $-13\pm$4   & $-14.5\pm$20 & $-16.5\pm$2   & $-12.5\pm$4 & $-9\pm$15 \\
C, km s$^{-1}$ kpc$^{-1}$         & $-2\pm$2      & $-5.5\pm$4  & $3.5\pm$20   & $2\pm$2       & $0.5\pm$4   & $21.5\pm$15 \\
K, km s$^{-1}$ kpc$^{-1}$         & $-1\pm$2      & $4.5\pm$4   & $-7.5\pm$20  & $-6\pm$2      & $4.5\pm$4   & $10.5\pm$15  \\
$l_{xy}^{\circ}$                  & $4\pm$4       & $11\pm$5    & $-19\pm$25   & $-3.5\pm$2    & $-1\pm$5    & $-23\pm$15 \\
$\Omega_{R0}$, km s$^{-1}$        & $-29\pm$5     & $-27\pm$5   & $-19\pm$36   & $-33\pm$4     & $-23\pm$6   & $-30\pm$28 \\
\hline
\end{tabular}
\]
\end{table*}
%¹¹¹¹¹¹¹¹¹¹¹¹¹¹¹¹¹¹¹¹¹¹¹¹¹¹¹¹¹¹¹¹¹¹¹¹¹¹¹¹¹¹¹¹¹¹¹¹¹¹¹¹¹¹¹¹¹¹¹¹¹¹¹¹¹

\section*{CONCLUSIONS}

Our study showed that the broadband multicolor photometry from the 2MASS and Tycho-2 catalogues
is enough to produce a pure (with an admixture of less than 1\%) sample of RGB stars complete
within 500 pc of the Sun, to classify the sample stars into types K, M, and C, suspected supergiants, and
S stars, to calculate the interstellar extinction, and to determine the photometric distances of K, M, and C
stars with an accuracy of 40\%. We showed that our classification is consistent with the old spectral one
and, in some cases, is even more accurate. 

Tycho-2 proper motions and PCRV radial velocities were used to analyze the stellar kinematics. The
decrease in the stellar density distribution with distance from the Galactic equator approximated by the
barometric law, contrary to the BMG, and the kinematic parameters calculated using the Ogorodnikov--Milne model
characterize the overwhelming majority of the selected K and M stars as disk stars with ages
of more than 3 Gyr. A small number of K and M stars are extremely young or, conversely, thick-disk
giants. The latter show a nonuniform distribution in the phase space of coordinates and velocities, arguing
against isothermality and full relaxation of the disk and for the theory of dynamical streams or superclusters.
The spatial distribution and kinematics of the selected C stars force us to consider them as
young (with ages $<2$ Gyr), fairly massive ($>2 M_{\odot}$) asymptotic branch giants probably associated with
the Gould Belt. The offset of the Sun above the Galactic equator was found from the distribution of
stars to be $13\pm2$ pc, which coincides with the previously obtained value for the RGC stars.

\section*{ACKNOWLEDGMENTS}

In this study, we used results from the Hipparcos and 2MASS (Two Micron All Sky Survey)
projects, the SIMBAD database (http://simbad.ustrasbg.fr/simbad/), and other resources of the Strasbourg
Data Center (France), http://cds.u-strasbg.fr/. The study was supported by the ``Origin and Evolution
of Stars and Galaxies'' Program of the Presidium of the Russian Academy of Sciences.

\begin{figure}[h]
\includegraphics{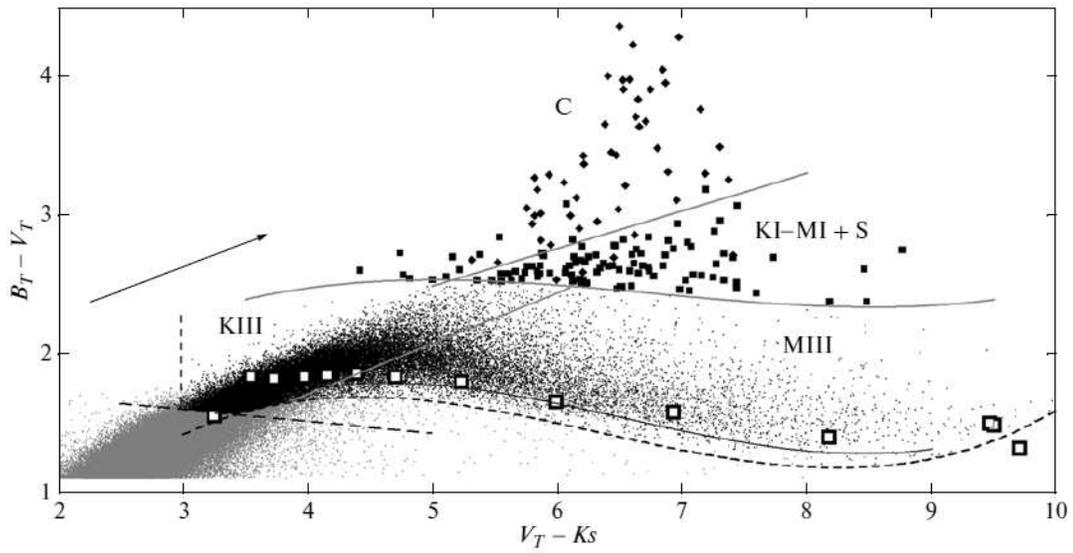}
\caption{$V_{T}-Ks$ -- $B_{T}-V_{T}$ diagram: the dashed lines separate the selected
RGB stars (black symbols) from the extraneous ones (gray symbols), the large open
squares mark the theoretical mean positions of solar-metallicity K3III--M10III stars,
the black solid curve indicates the line of normal colors for MIII stars,
and the gray lines divide the selected stars into four categories: KIII, MIII, S/supergiants, and type C.
}
\label{vkbv}
\end{figure}

\begin{figure}[h]
\includegraphics{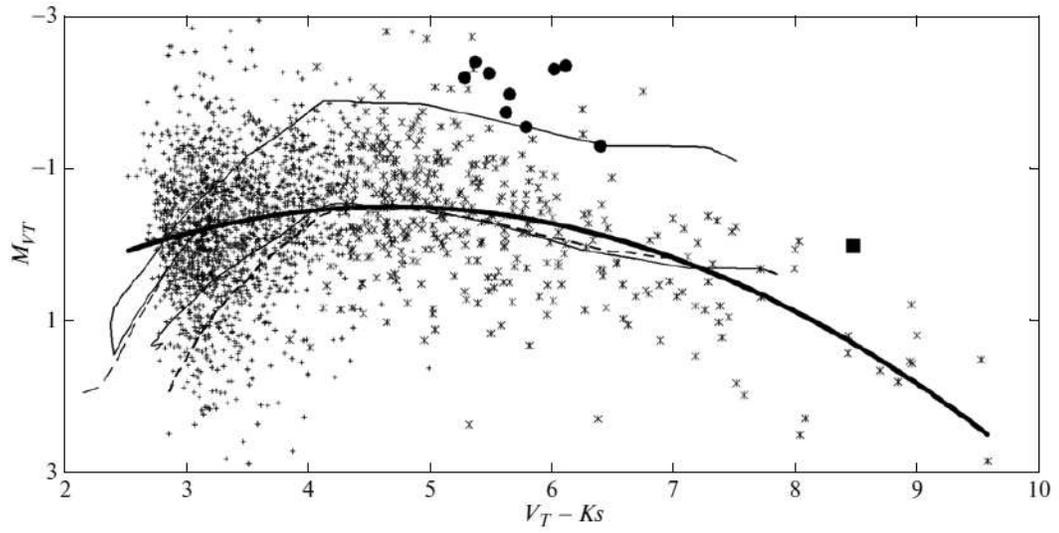}
\caption{$M_{VT}$ versus $V_T-Ks$ for the selected K (crosses), M (snowflakes),
C (circles), and S (square) stars. The thick curve
indicates the adopted calibration. The isochrones for solar-metallicity stars
are indicated by the thin solid (asymptotic giant branch) and dashed (RGB) curves;
the upper solid and dashed lines for an age of 10$^9$ yr and the lower one
for 10$^{10}$ yr.
}
\label{vkmv}
\end{figure}

\begin{figure}[h]
\includegraphics{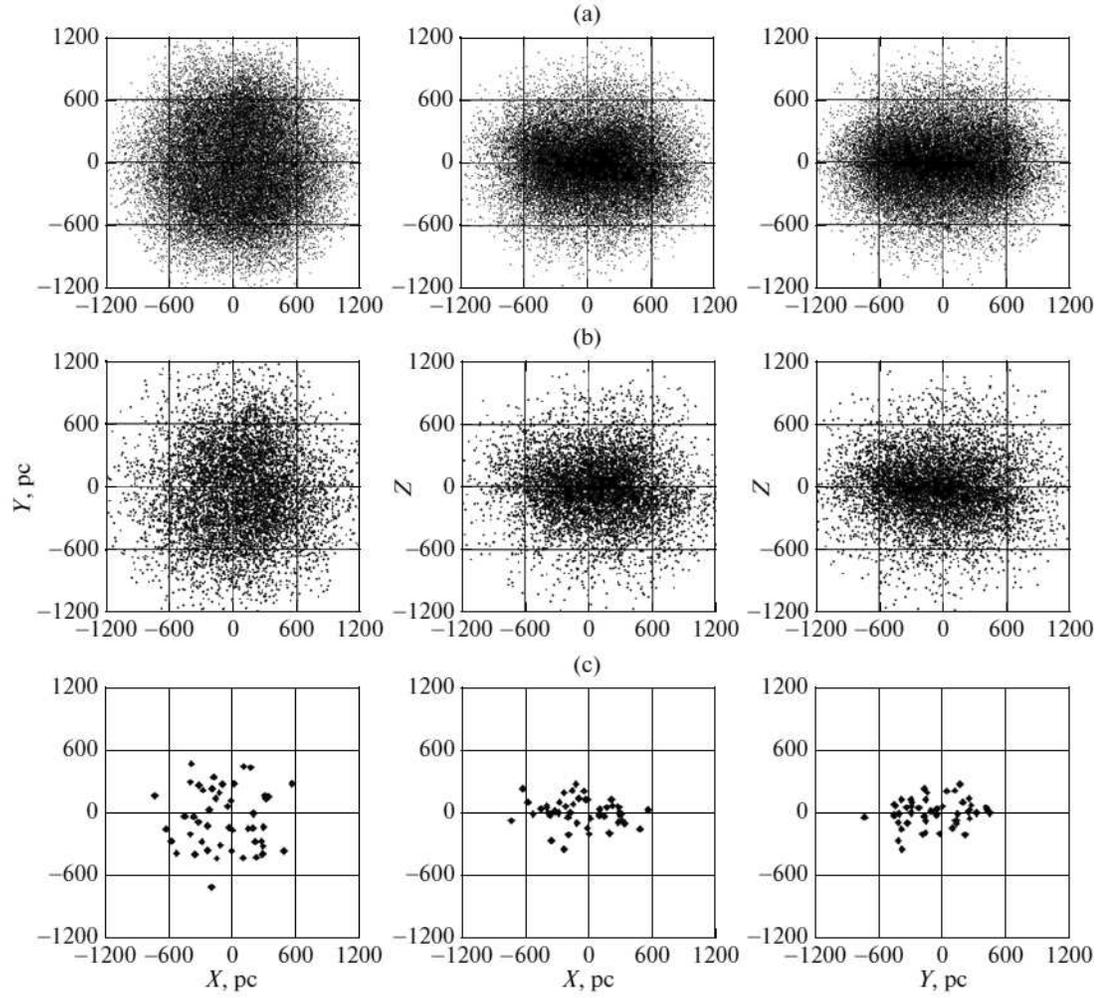}
\caption{Distribution of the selected stars in projection onto the $XY$, $XZ$, and
$YZ$ planes: (a) type K, (b) type M, and (c) type C.
}
\label{xyz}
\end{figure}

\begin{figure}[h]
\includegraphics{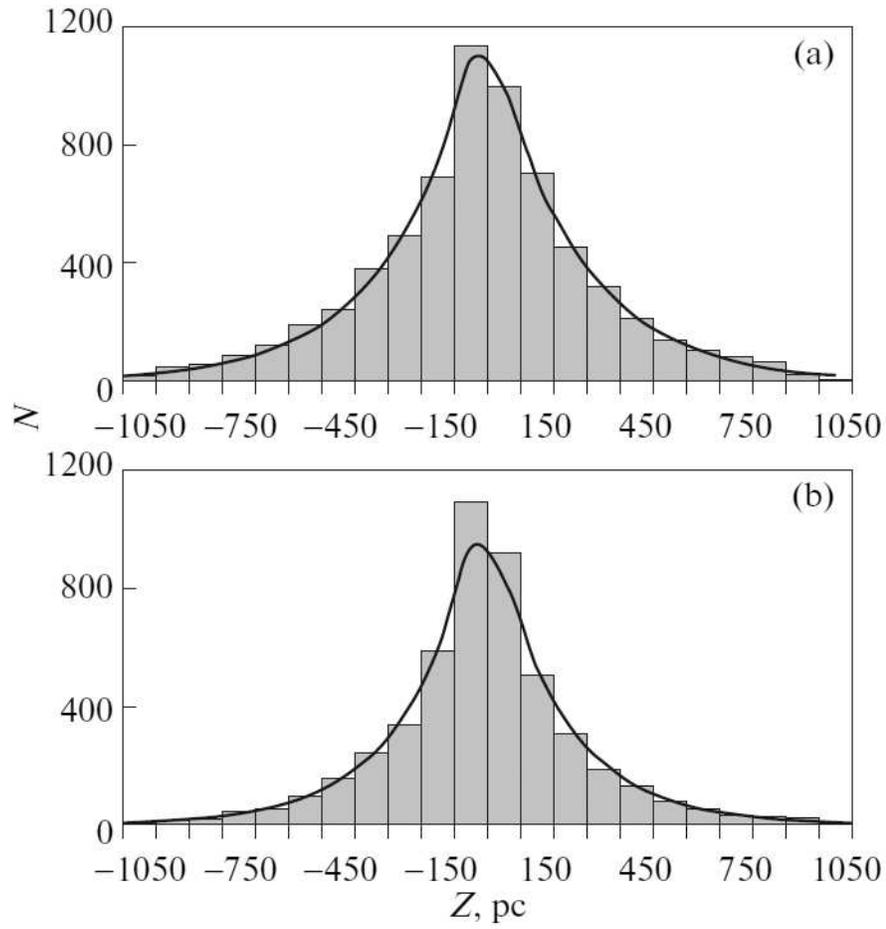}
\caption{Distribution of the selected (a) K and (b) M stars along the $Z$ axis
in the vertical cylinder within 400 pc of the Sun; the curves indicate the fit by the barometric law.
}
\label{zN}
\end{figure}

\begin{figure}[h]
\includegraphics{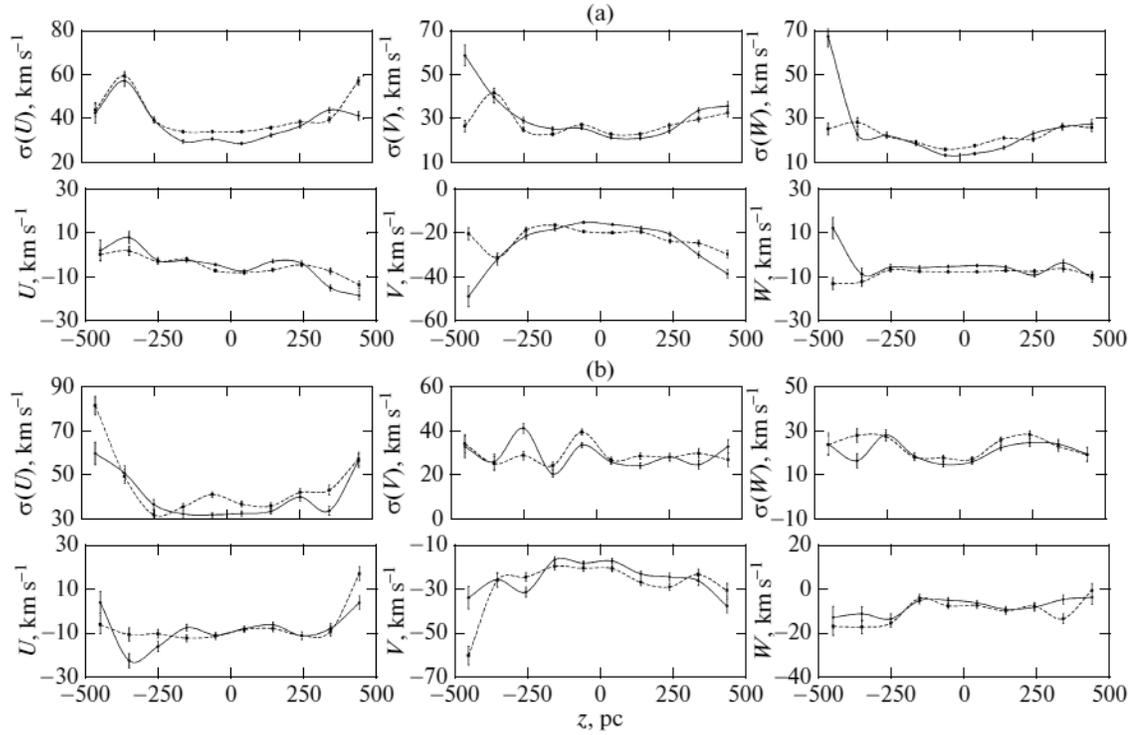}
\caption{Mean velocity and velocity dispersion in km s$^{-1}$ versus $Z$ coordinate in pc in the vertical
cylinder elongated along the $Z$ axis within 500 pc of the Sun: (a) type K, (b) type M. The results from
photometric distances and Hipparcos parallaxes are indicated by the solid and dashed lines, respectively.
}
\label{zuvw}
\end{figure}

\end{document}